\documentstyle[sprocl]{article}

\input{psfig}
\def\Journal#1#2#3#4{{#1} {\bf #2}, #3 (#4)}

\def\PLB{{\em Phys. Lett.}  B}
\def\PRL{\em Phys. Rev. Lett.}
\def\PRD{{\em Phys. Rev.} D}

\def\be{\begin{equation}}
\def\ee{\end{equation}}
\def\bea{\begin{eqnarray}}
\def\eea{\end{eqnarray}}

\begin{document}

\title{Dynamics of Strangeness Production and Strange Matter Formation
\footnote[1]{Supported by GSI, BMBF, DFG}}

\author{C.~Spieles, M.~Bleicher, L.~Gerland, H.~St\"ocker}

\address{Institut f\"ur
Theoretische Physik,  J.~W.~Goethe-Universit\"at,\\
D-60054 Frankfurt am Main, Germany}

\author{C.~Greiner}

\address{Institut f\"ur Theoretische Physik,
J.~Liebig-Universit\"at,\\
D-35392 Giessen, Germany}


\maketitle

\section{Introduction}
We want to draw the attention to the dynamics of a (finite) hadronizing
quark matter drop. Strange and antistrange quarks do not hadronize
at the same time for a baryon-rich system\cite{CG1}. Both the hadronic and the quark matter phases enter the strange sector
$f_s\neq 0$ of the phase
diagram almost immediately, which 
has up  to now been neglected in almost all calculations of the time
evolution of the system.  Therefore it seems questionable, whether final
particle yields reflect the actual thermodynamic properties of the system at
a certain stage of the evolution. We put special interest on the possible
formation of exotic states, namely strangelets (multistrange quark
clusters).
They may exist as (meta-)stable exotic isomers of nuclear matter \cite{Bod71}.
It was speculated  that strange matter might
exist also as metastable exotic multi-strange (baryonic) 
objects (MEMO's \cite{Sch92}).
The possible creation --- in heavy ion collisions ---
 of long-lived remnants of the quark-gluon-plasma, cooled and charged
up with strangeness by the emission of pions and kaons, 
was proposed in \cite{CG1,Liu84,CG2}.
Strangelets can serve as signatures for the
creation of a quark gluon plasma.
Currently, both at the BNL-AGS and at the CERN-SPS experiments are carried 
out to search for MEMO's and strangelets, e.~g. by the E864, E878 and the NA52
collaborations\cite{qm95,str95}.

\section{The model}
We adopt a model \cite{CG2} for the hadronization and space-time evolution
of quark matter droplet. We assume
 a first order phase transition of the QGP to hadron gas.
The expansion of the QGP droplet is described in a hybrid-like model,
which takes into account equilibrium as well as nonequilibrium features
of the process by the following two crucial, yet oversimplifying
(and to some extent controversial) assumptions:
(1) the plasma sphere is permantently surrounded by a thin layer of hadron gas,
with which it stays in perfect equilibrium (Gibbs conditions) during the 
whole evolution; in particular the strangeness degree of freedom stays in 
chemical equilibrium because the complete hadronic particle production is 
driven by the plasma phase.  
(2) The nonequilibrium radiation is incorporated by 
a time dependent freeze-out of hadrons from the outer layers of the 
hadron phase surrounding the QGP droplet.
During the expansion, the volume increase of the system thus
competes with the decrease due to the freeze--out.
The global properties like (decreasing) $S/A$ and (increasing) $f_s$ 
 of the remaining two-phase system then change in time
according to the following differential equations for the baryon number, the
entropy, and the net strangeness number of the total system:
\begin{eqnarray}\label{eq1} 
\frac{d}{dt}A^{tot}  & = & -\Gamma \, A^{HG}  \nonumber \\
\frac{d}{dt}S^{tot}  & = & -\Gamma \, S^{HG} \\
\frac{d}{dt}(N_s - N_{\overline{s}})^{tot}  & =  & -\Gamma \,
(N_s - N_{\overline{s}})^{HG} \, \, \, , \nonumber
\end{eqnarray}
where $\Gamma = \frac{1}{A^{HG}}
\left( \frac{\Delta A^{HG}}{\Delta t} \right) _{ev}$ is the effective
(`universal') rate of particles (of converted hadron gas volume)
evaporated from the hadron phase.
The equation of state consists of the bag model for the
quark gluon plasma and a
mixture of relativistic Bose--Einstein and Fermi--Dirac gases of well
established strange and non--strange hadrons up to 2~GeV 
in Hagedorn's eigenvolume correction for the
hadron matter \cite{CG1}. 
Thus, one solves simultaneously
the equations of motion (\ref{eq1}) and the Gibbs phase equilibrium
conditions for the intrinsic variables, i.e. the chemical potentials and the
temperature, as functions of time.

\begin{figure}[t]
\vspace*{-0.6cm}
\begin{minipage}[b]{2in}
\vskip 0mm
\hspace*{-0.2cm}
\psfig{figure=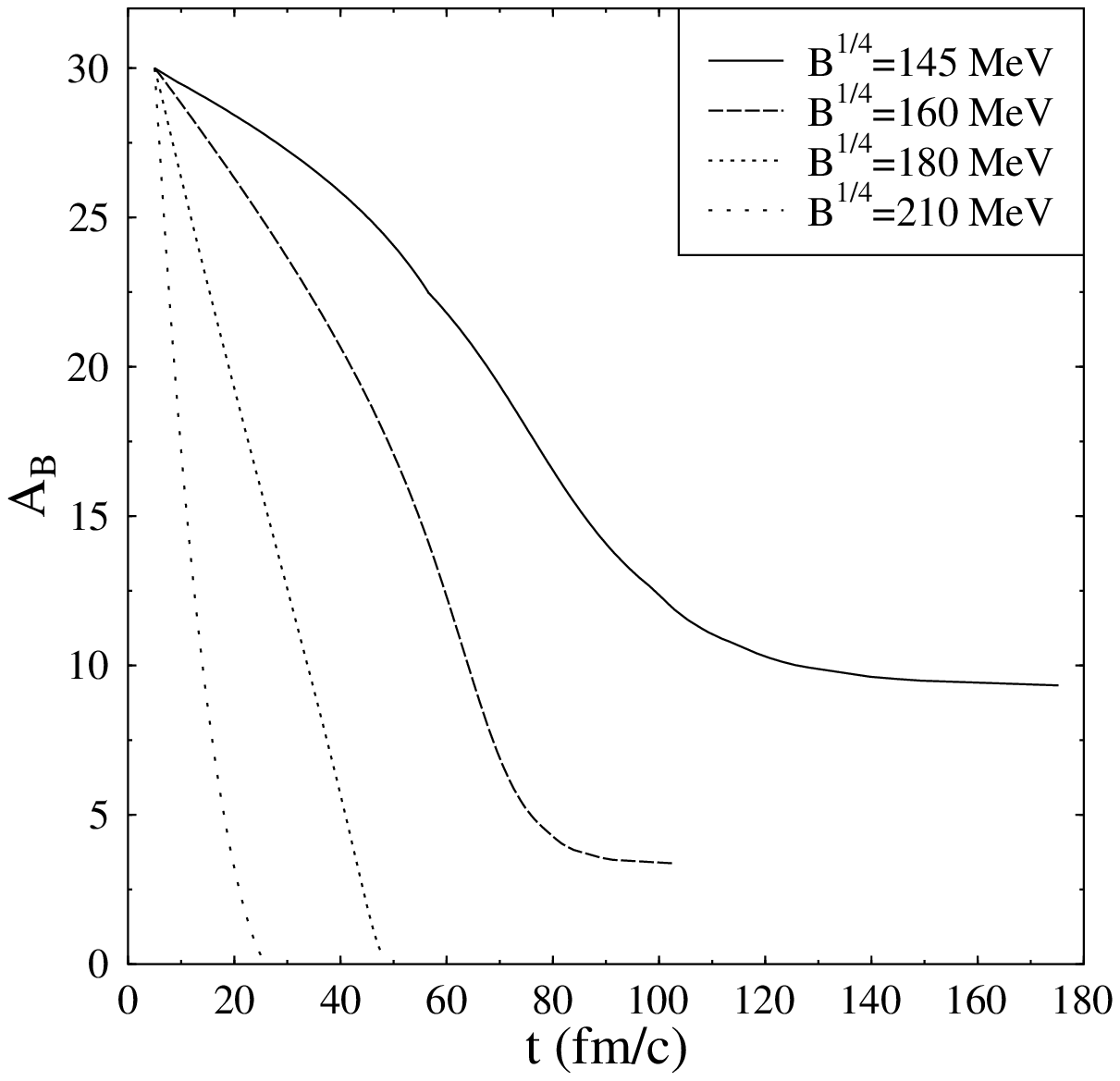,height=2.5in}
\vskip 2mm
\vspace{0.0cm}
\vspace*{-0.45cm}
\end{minipage}
\hfill
\begin{minipage}[b]{2in}
\vskip 0mm
\hspace*{-1.2cm}
\psfig{figure=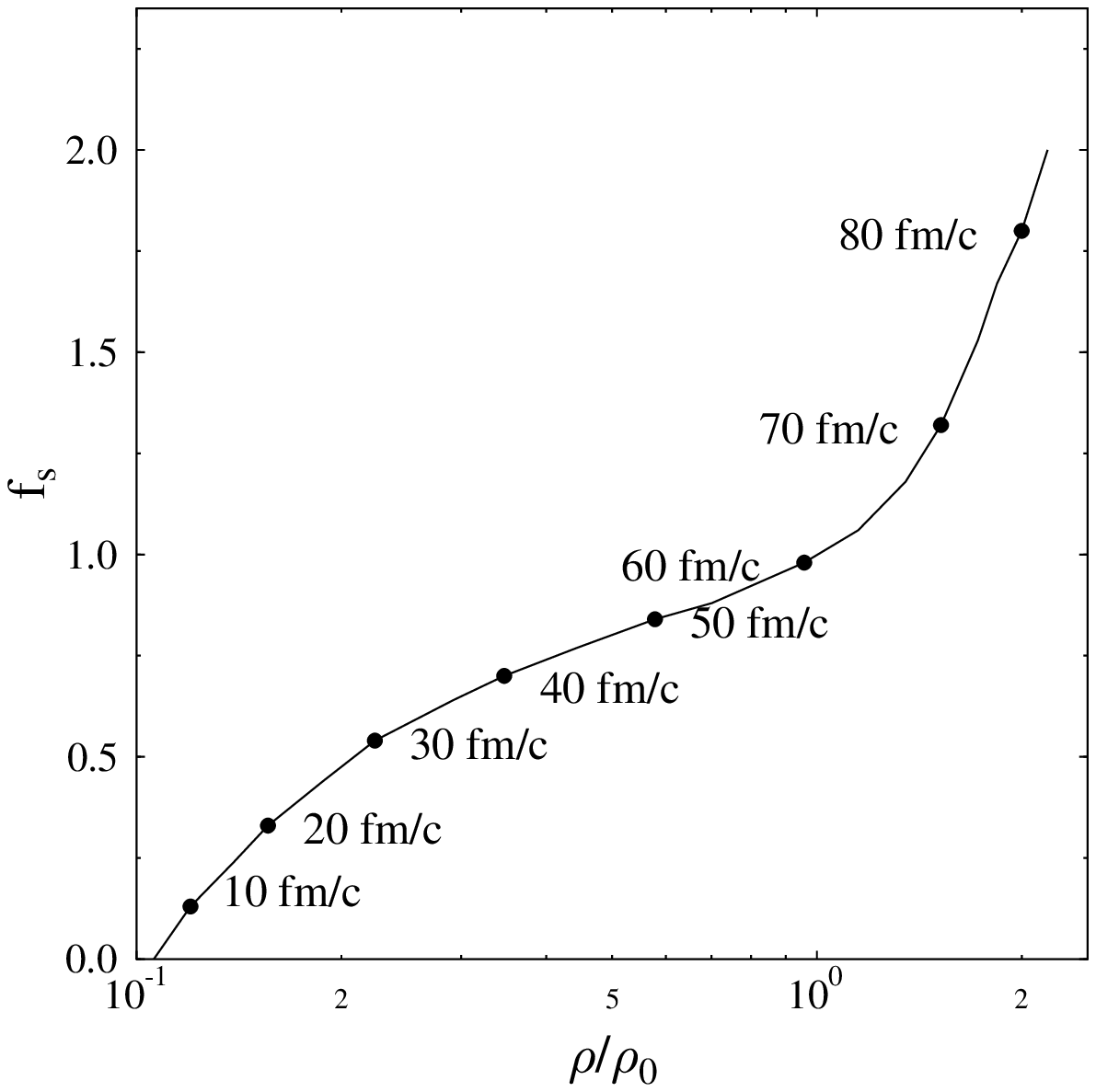,height=2.5in}
\vskip 2mm
\vspace{0.0cm}
\vspace*{-0.45cm}
\end{minipage}
\caption{Time evolution of the baryon number for 
a QGP droplet with $A_{\rm B}^{\rm init}=30$, $S/A^{\rm init}=200$, 
$f_s^{\rm init}=0.7$ 
and different bag constants (left).
Evolution of a QGP droplet with baryon number $A_{\rm B}^{\rm
init}=30$ for $S/A^{\rm init}=200$ and
$f_s^{\rm init}=0$. The bag constant is $B^{1/4}=160$~MeV. Shown is the
baryon density and the corresponding strangeness fraction (right).
\label{abdampf}}
\vspace*{-0.35cm}
\end{figure}

\section{Strangelet distillation at low $\mu/T$}
In \cite{Sp96} it was shown that large
local net-baryon and net-strangeness fluctuations as well as 
a small but finite amount of stopping can occur at RHIC and LHC.
This can provide 
suitable initial conditions for the possible creation of strange matter 
in colliders. 
A phase transition (e.~g. a chiral one) can 
further increase the strange matter formation probability. 
In \cite{Sp96} it was further demonstrated with the 
present model that the high initial
entropies per baryon do not hinder the distillation of strangelets, however,
they require more time for the evaporation and cooling process.

\begin{figure}[t]
\vspace*{-0.7cm}
\begin{minipage}[b]{2in}
\vskip 0mm
\vspace{0.5cm}
\hspace*{-1.cm}
\psfig{figure=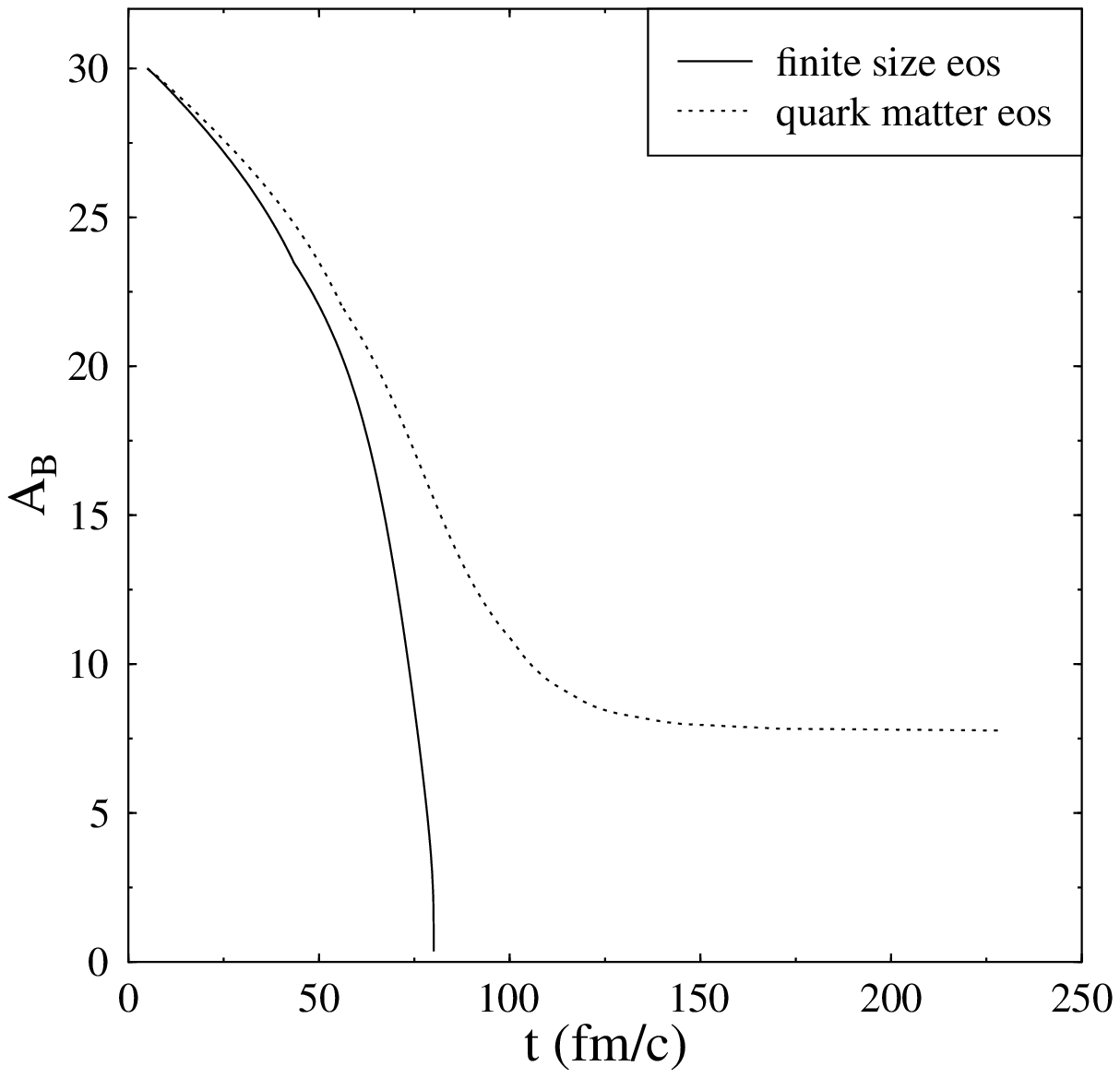,height=2.5in}
\vskip 2mm
\vspace{0.0cm}
\vspace*{-1.1cm}
\end{minipage}
\hfill
\begin{minipage}[b]{2in}
\vskip 0mm
\vspace{0.5cm}
\hspace*{-1.5cm}
\psfig{figure=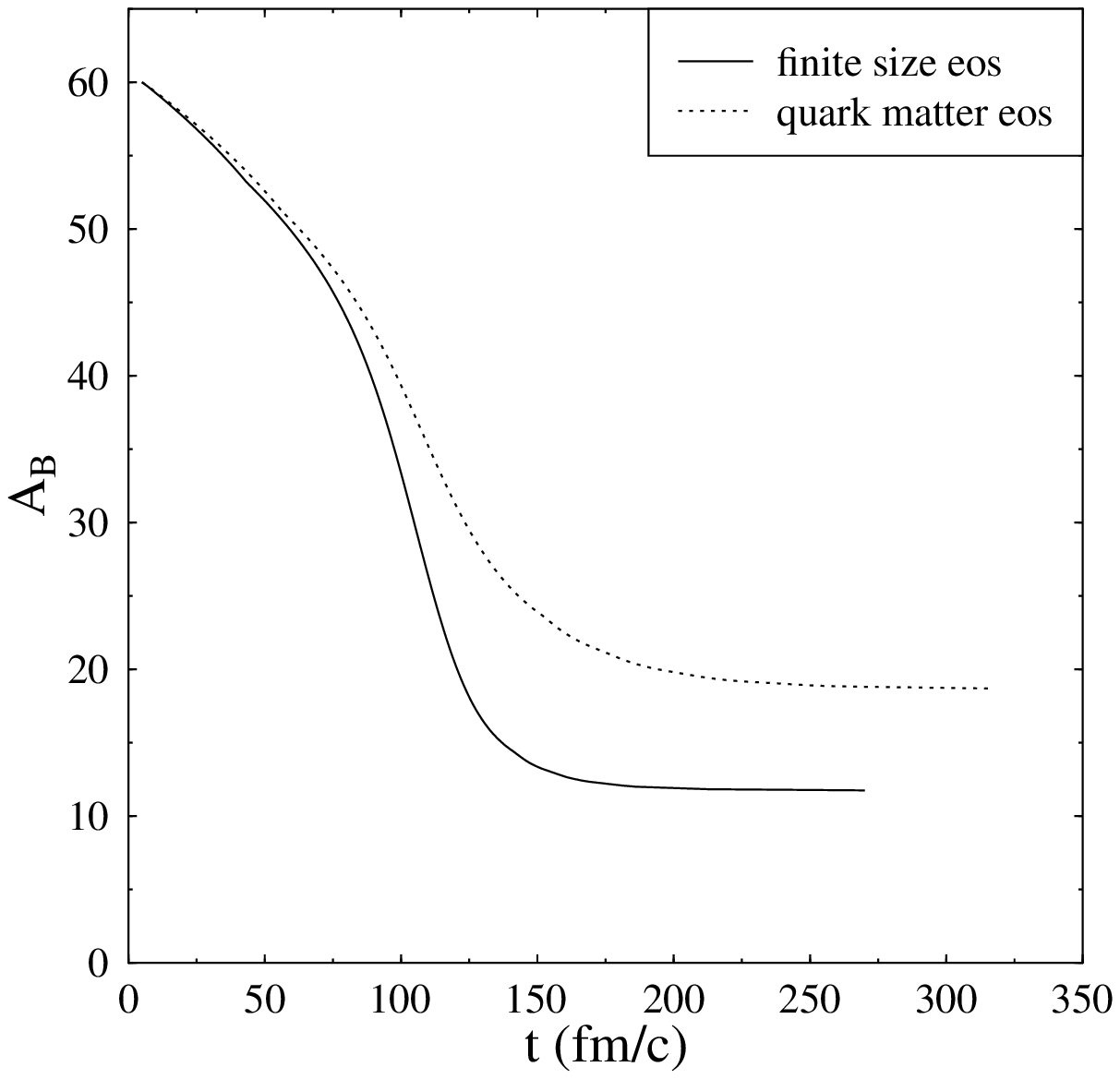,height=2.5in}
\vskip 2mm
\vspace{0.0cm}
\vspace*{-1.1cm}
\end{minipage}
\caption{
Time evolution of the net baryon number of a QGP droplet,
calculated with (full line) and without (dashed line) finite size
corrections to the quark matter equation of state.
The initial conditions are 
$f_s^{\rm init}=0$ and $S/A^{\rm init}=200$.
The bag constant is $B^{1/4}=145$~MeV. \label{finsize}}
\vspace*{-0.45cm}
\end{figure}

\begin{figure}
\vspace*{-0.6cm}
\begin{minipage}[b]{2in}
\vskip 0mm
\vspace{0.5cm}
\hspace*{-0.1cm}
\psfig{figure=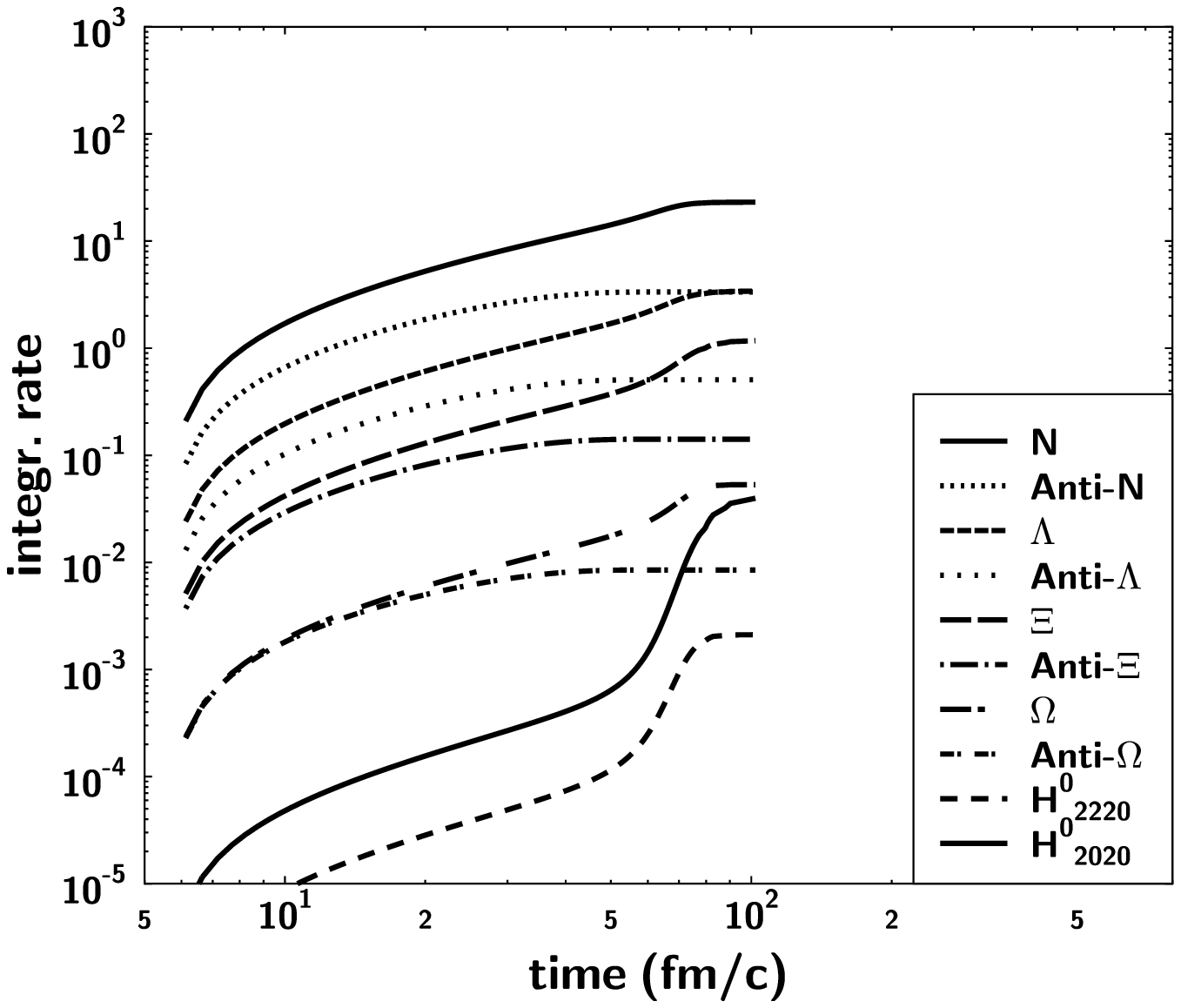,height=2in}
\vskip 2mm
\vspace{0.0cm}
\vspace*{-0.3cm}
\end{minipage}
\hfill
\begin{minipage}[b]{2in}
\vskip 0mm
\vspace{0.5cm}
\hspace*{-1.3cm}
\psfig{figure=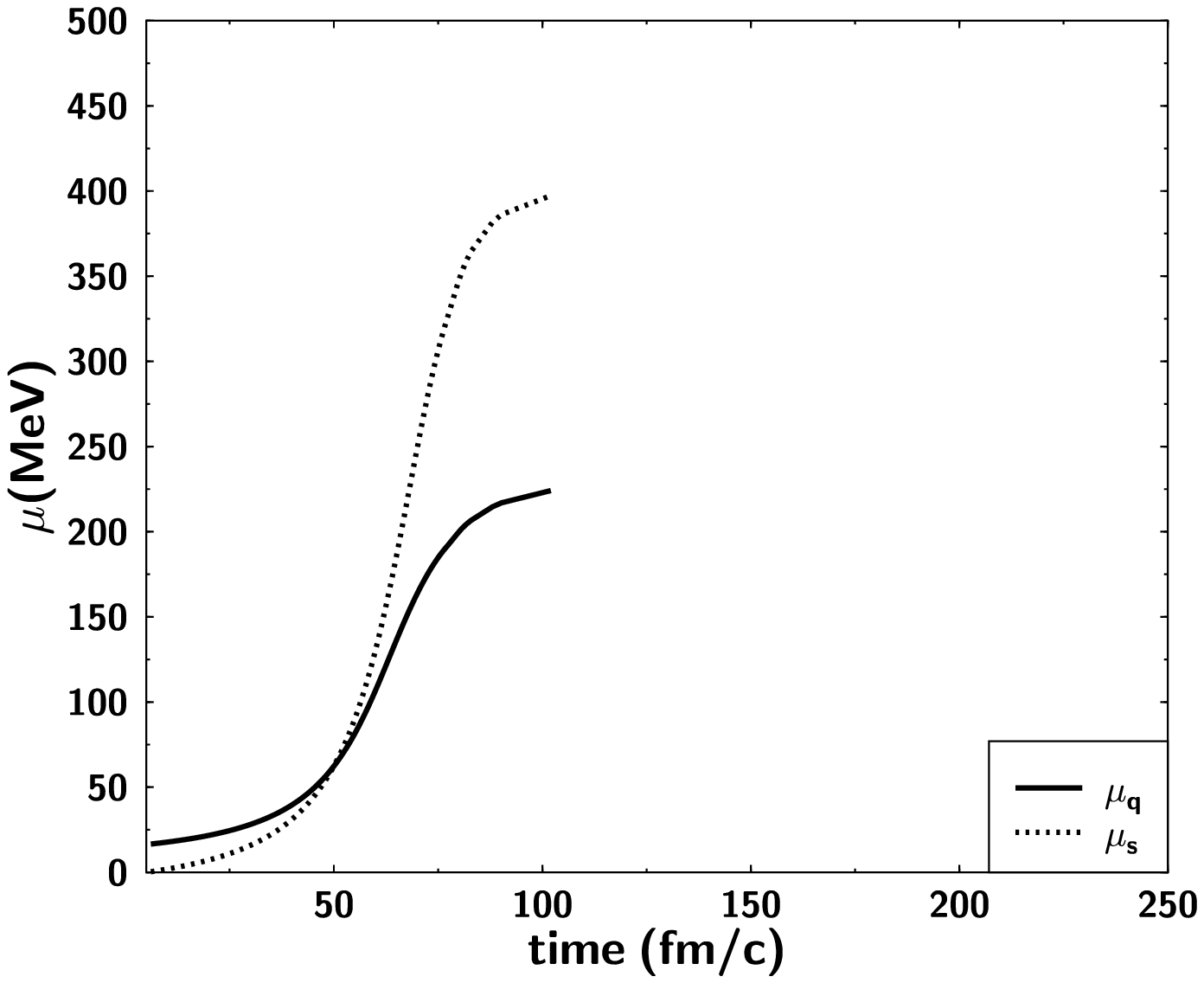,height=2in}
\vskip 2mm
\vspace{0.0cm}
\vspace*{-0.3cm}
\end{minipage}
\caption{
Integrated rates of particles, evaporated out of a hadronizing QGP droplet
as functions of time (left) and the corresponding (strange) quark chemical
potential (right).
The initial conditions are 
$f_s^{\rm init}=0$, $S/A^{\rm init}=200$ and $A_{\rm B}^{\rm
init}=30$. The bag constant is $B^{1/4}=160$~MeV, the mass of the
$H^0$ is varied between $m=2020$~MeV and $m=2220$~MeV.\label{rates1}}
\vspace*{-0.3cm}
\end{figure}

\begin{figure}
\vspace*{-0.6cm}
\begin{minipage}[b]{2in}
\vskip 0mm
\vspace{0.5cm}
\hspace*{-0.1cm}
\psfig{figure=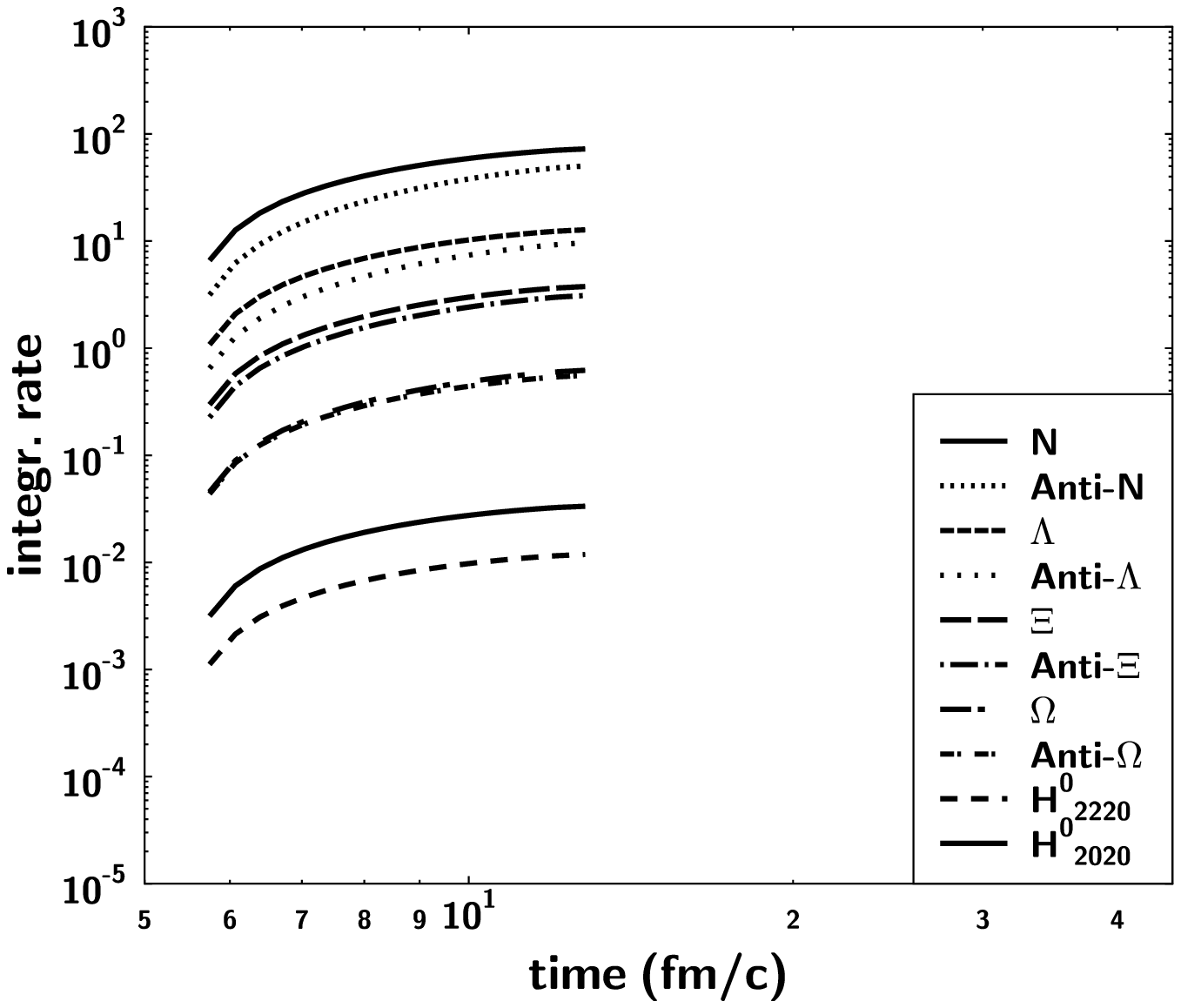,height=2.1in}
\vskip 2mm
\vspace{0.0cm}
\vspace*{-0.55cm}
\end{minipage}
\hfill
\begin{minipage}[b]{2in}
\vskip 0mm
\vspace{0.5cm}
\hspace*{-1.3cm}
\psfig{figure=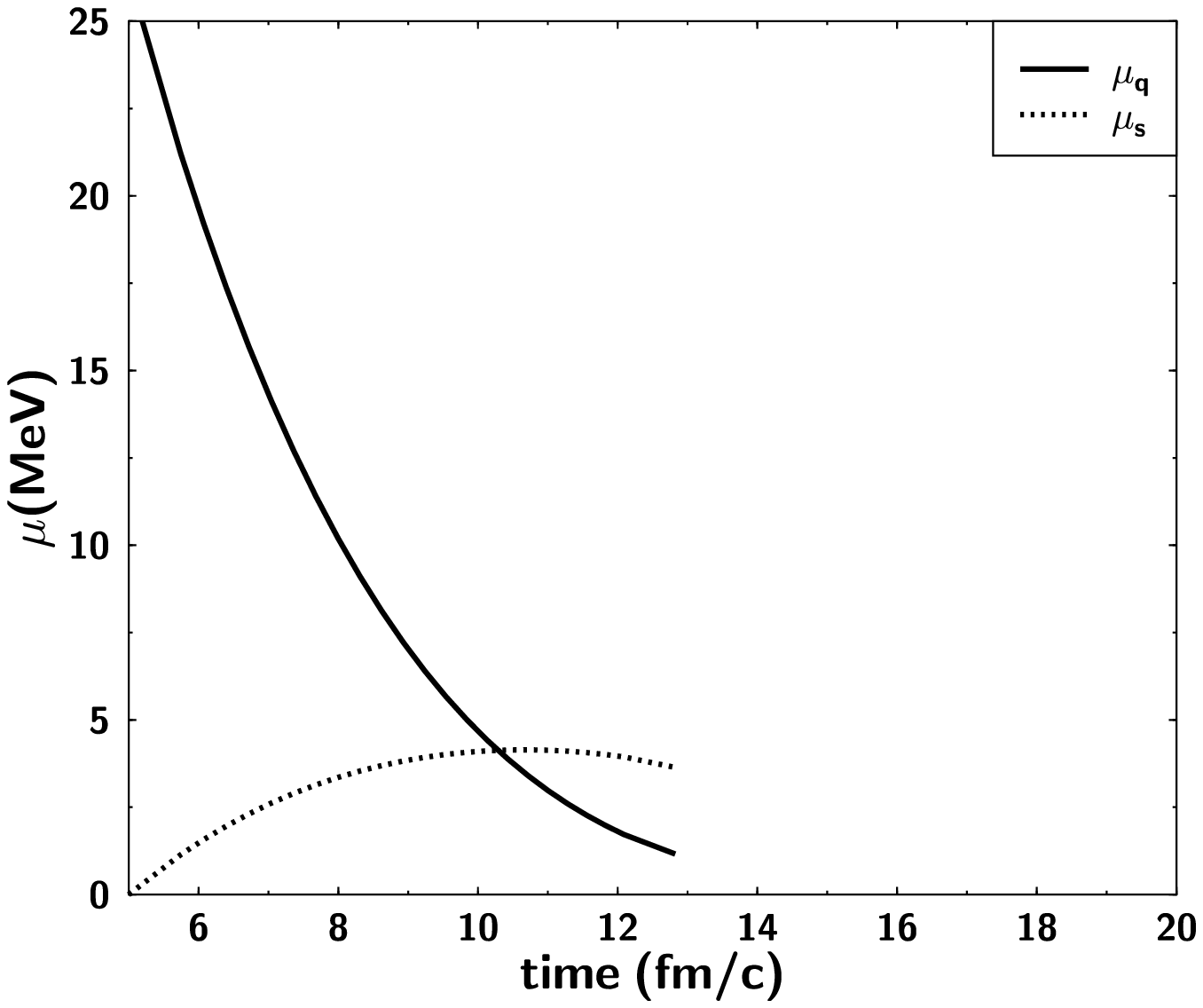,height=2.1in}
\vskip 2mm
\vspace{0.0cm}
\vspace*{-0.55cm}
\end{minipage}
\caption{
Integrated rates of particles, evaporated out of a hadronizing QGP droplet
as functions of time (left) and the corresponding (strange) quark chemical
potential (right).
The initial conditions are 
$f_s^{\rm init}=0$, $S/A^{\rm init}=200$ and $A_{\rm B}^{\rm
init}=30$. The bag constant is $B^{1/4}=250$~MeV, the mass of the
$H^0$ is varied between $m=2020$~MeV and $m=2220$~MeV.\label{rates2}}
\vspace*{-0.6cm}
\end{figure}

Fig.~\ref{abdampf} (left) shows the time evolution of the baryon number for 
 $S/A^{\rm init}=200$ and $f_s^{\rm init}=0.7$ for  various
bag constants.
For $B^{1/4}<180$~MeV
a cold
strangelet emerges from the expansion and evaporation process, while
the droplet completely hadronizes for bag constants $B^{1/4}\ge 180$~MeV 
(for $B^{1/4}=210 $ MeV hadronization proceeds
without any significant 
cooling of the quark phase, although the specific entropy $S/A$
decreases by a factor of 2 from 200 to only 100).
The strangeness
separation works also in these cases, and leads to 
large final values of the
net strangeness content,
$f_s \stackrel{>}{\sim } 1.5-2$. However, then the volume of the drop 
becomes small, it decays and the
strange quarks hadronize
into $\Lambda $-particles and other strange hadrons.
For even higher bag constants $B^{1/4}\approx 250$~MeV neither the baryon
concentration effect nor strangeness distillery occurs (Fig.~\ref{rates2}). 

Fig.~\ref{abdampf} (right) shows the evolution of the two-phase 
system for $S/A^{\rm
init}=200$, $f_s^{\rm init}=0$ and for a bag constant $B^{1/4}=160$~MeV 
in the plane of the strangeness fraction vs. the baryon density.
The baryon density increases by more than one order of magnitude!
Correspondingly, the chemical potential rises as drastically
during the evolution, namely from $\mu^i=16$~MeV to
 $\mu^f>200$~MeV. 
The strangeness
separation mechanism drives the chemical potential of the strange quarks 
from $\mu^i_s=0$ up to $\mu^f_s\approx 400$~MeV.
Thus, the thermodynamical and
chemical properties during the time evolution 
are quite different from
 the initial conditions of the system. 

Fig.~\ref{abdampf} illustrates
the increase of the baryon density in the plasma droplet
as an inherent feature of the dynamics of the phase
transition (cf. \cite{Wi84}).
The origin of this result lies in
 the fact that the baryon number in the
quark--gluon phase is carried by quarks with $m_{\rm q}\ll T_{\rm C}$, while
the baryon density in the hadron phase is suppressed by a Boltzmann factor
$\exp (-m_{\rm baryon}/T_{\rm C})$ with $m_{\rm baryon}\gg T_{\rm C}$.
Mainly mesons (pions and kaons) are created 
in the hadronic phase. 
More relative entropy $S/A$ than baryon number is carried away in the 
hadronization and
evaporation process\cite{CG2}, i.e.
$(S/A)^{HG} \gg (S/A)^{QGP}$. Ultimately, whether $(S/A)^{HG}$ is larger or
smaller than $(S/A)^{QGP}$ at finite, nonvanishing chemical potentials might
theoretically only be proven rigorously by lattice gauge calculations in the
future. However, model equations of state do suggest such a behaviour,
which would open such intriguing possibilities as baryon inhomogenities in
ultrarelativistic heavy ion collisions as well as in the early universe.

\section{Finite size effects}

The bag model equation of state for infinite quark matter is certainly a
very rough approximation.
Regarding finite size effects the leading correction to the quark matter
equation of state is the curvature term. For massless quarks the
volume term of the gandcanonical potential suffers the following
modification (including the gluon contribution)\cite{Mad93}:
\begin{equation}
\Omega_{\rm C}
=(\frac{1}{8\pi^2}\mu_{\rm q}^2+\frac{11}{72}T^2)C
\end{equation}
with $C=8\pi R$ being the curvature of the spherical bag surface. From this
one can easily derive all thermodynamic quantities and study the evolution
of the two-phase system QGP/hadron gas in the above described model.
It shows that even in the case of a favourable bag constant
$B^{1/4}=145$~MeV a quark blob with an 
initial net baryon number of  $A_{\rm B}^{\rm
init}=30$ will completely hadronize --- in contrast to the calculation with
the unmodified equation
of state (Fig.~\ref{finsize}). Of course, the difference between the
dynamics according to the two equations of state is reduced for larger
systems.
Still, it can be speculated that shell effects may allow for the
formation of rather small strangelets which are stable. Moreover, the
introduction of a more realistic hadronic equation of state (e.~g. with the
help of a relativistic mean field theory including adequate interactions for
strange hadrons \cite{Sch92}) might modify this pessimistic picture again.

\section{Particle rates from the hadronizing plasma}
Enhanced production of strange particles in relativistic nuclear collisions
has received much attention recently\cite{qm95,str95}. In particular thermal
models have been developed and applied\cite{Ra95,PBM} to explain (strange)
particle yields and to extract the characteristic thermodynamic properties
of the system (a few macroscopic parameters) from them.
In our model the picture of a sudden hadronization which is supposed in
these studies is only one possible outcome. Under more general assumptions
the observed particle rates have to be put in relation to the whole time
evolution of the system.

The integrated particle rates and the quark chemical potentials as functions
of time have been calculated for two different scenarios: In
Fig.~\ref{rates1} the
results are plotted for a bag constant of $B^{1/4}=160$~MeV which is
favorable for the strangeness distillation. In Fig.~\ref{rates2} a very high bag
constant of $B^{1/4}=250$~MeV is used. This results in a very rapid (and
complete) hadronization without significant cooling.
Obviously, in the first case the particle rates reflect the massive changes
of the chemical potentials during the evolution (which is the result of the
strangeness distillation process). Note that e.~g. the $\Lambda$'s are
emitted mostly at the late stage, whereas the $\bar \Lambda$'s stem almost
exclusively from the early stage. The $\bar \Lambda/\Lambda$ ratio is
therefore not a meaningful quantity (if one takes it naively), since the
two yields represent
different sources! For the other choice of the bag constant the present
model renders more or less the picture which is claimed by 'static' thermal
models: the plasma fireball decomposes very fast into hadrons (watch the
different time scales of Figs.~\ref{rates1} and \ref{rates2}) and the quark chemical
potentials stay low compared to the temperature.
Time dependent rates of the hypothetic $H^0$ Dibaryon are also shown in
Figs.~\ref{rates1} and \ref{rates2}. This particle is introduced to the hadronic
resonance gas with its appropriate quantum numbers and two different assumed
masses. It appears that the distillation mechanism gives rise to $H^0$
yields of the same order as the $\bar \Omega$'s (Fig.~\ref{rates1}) if the mass is
$m_{H^0}\approx 2020$~MeV. For the high bag constant the $H^0$ yields are
much more suppressed as compared to the strange (anti-)baryons. The
absolute yields of the $H^0$ do not change much, since the system emits the
particles at significantly higher temperature (due to the high bag
constant).

\begin{figure}
\vspace*{-0.3cm}
\begin{minipage}[b]{2in}
\vskip 0mm
\vspace{0.5cm}
\psfig{figure=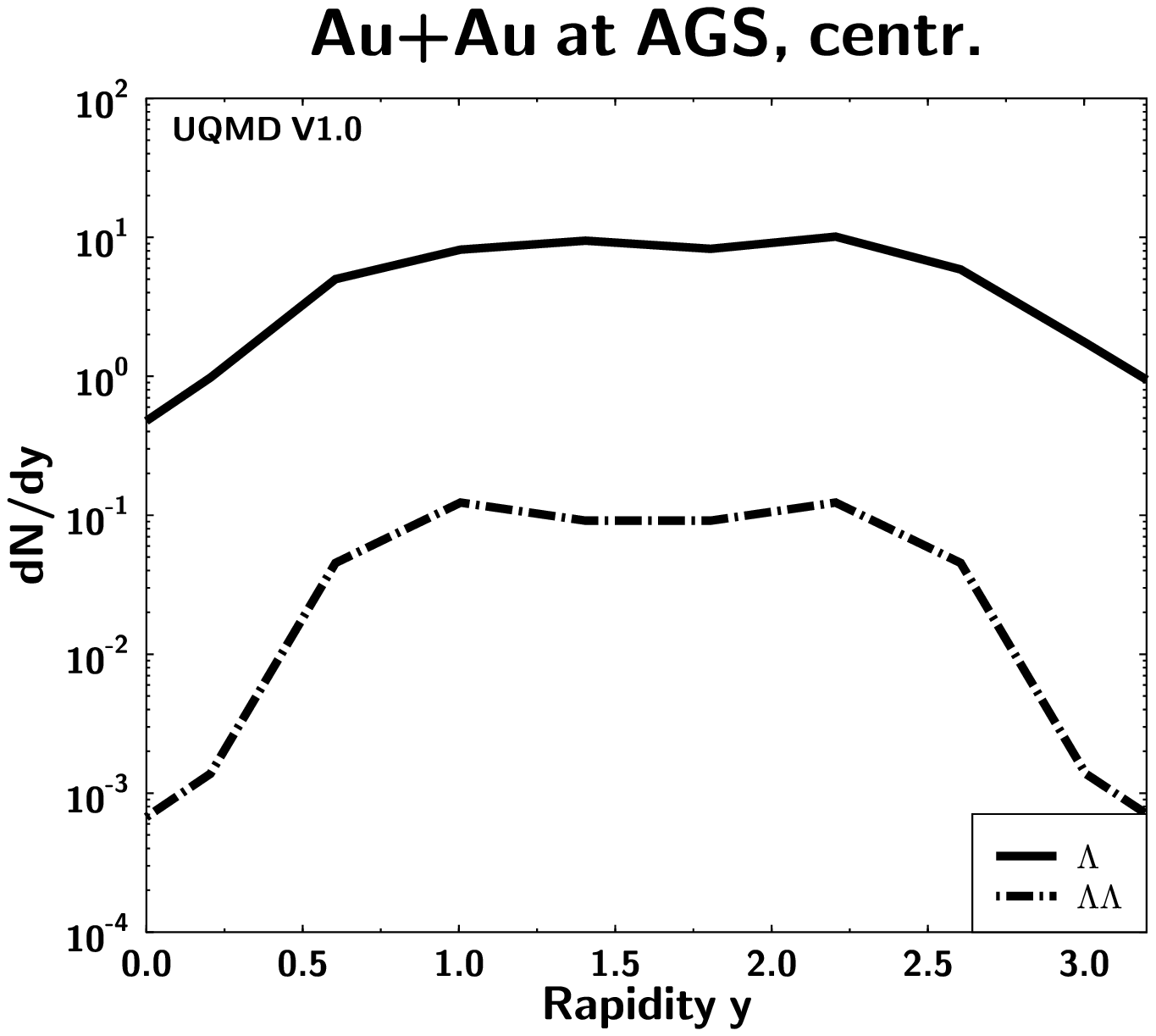,height=2in}
\vskip 2mm
\vspace{0.0cm}
\vspace*{-0.3cm}
\end{minipage}
\hfill
\begin{minipage}[b]{2in}
\vskip 0mm
\vspace{0.5cm}
\hspace*{-1.2cm}
\psfig{figure=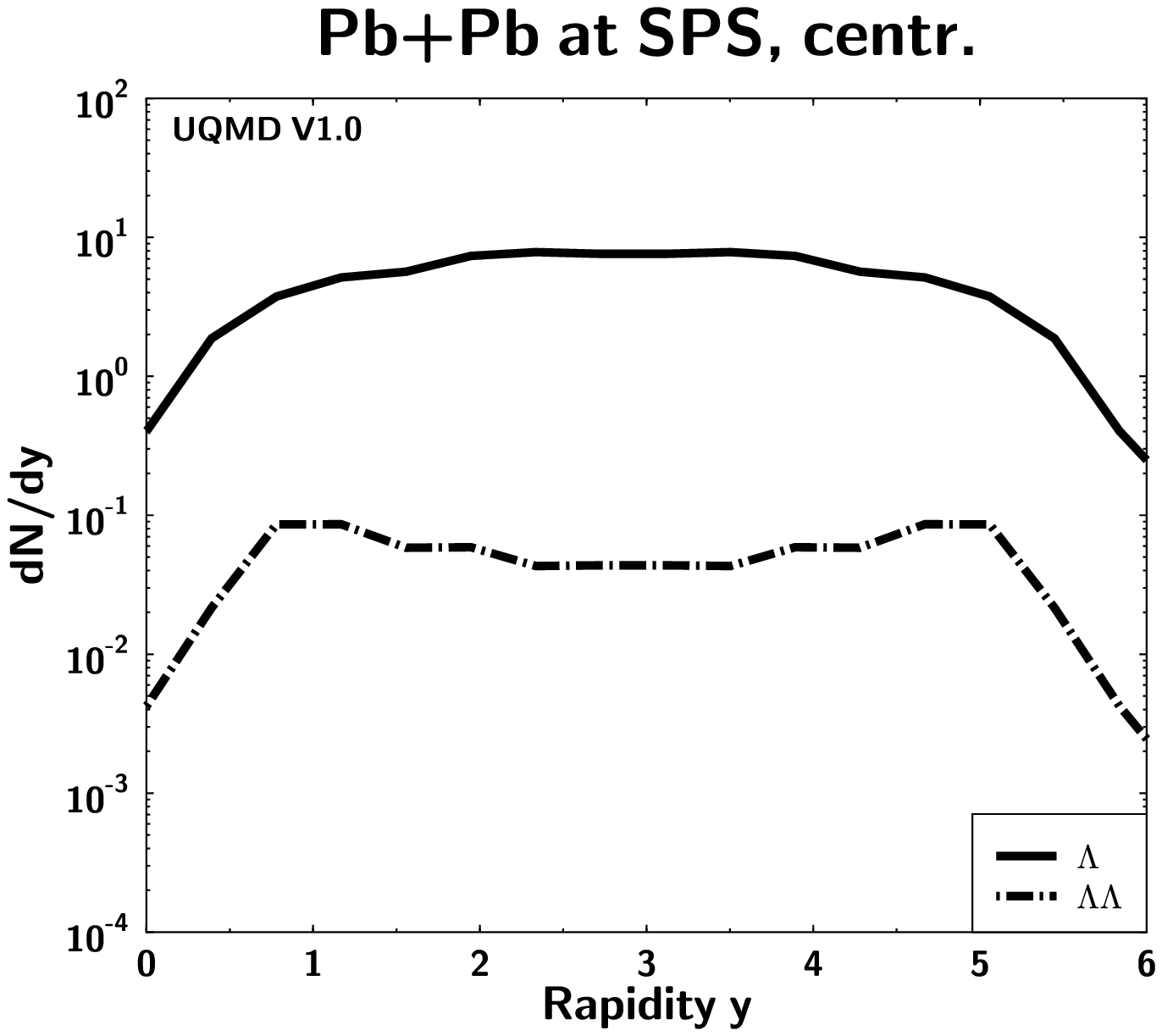,height=2in}
\vskip 2mm
\vspace{0.0cm}
\vspace*{-0.3cm}
\end{minipage}
\caption{Rapidity distributions of hyperons and $\Lambda-\Lambda$-clusters
calculated with URQMD$1.0\beta$ plus a clustering procedure according to
the Wigner-function method. Shown are the spectra for central collisions of
Au+Au at $10.7$~GeV (left) and Pb+Pb at 160~GeV (right).
\label{yclust}}
\end{figure}

\section{Hyper--cluster formation in a microscopic model}
We now apply the Ultrarelativistic Quantum Molecular Dynamics 1.0$\beta$
\cite{uqmd,bleicher}, 
a semiclassical transport model, to calculate the abundances of strange
baryon-clusters in relativistic heavy ion collisions. The model is based on
classical propagation of hadrons and stochastic scattering ($s$ channel
excitation of baryonic and mesonic resonances/strings, $t$ channel
excitation, deexcitation and decay). 
In order to extract hyper-cluster formation probabilities the $\Lambda$ pair
phase space after strong freeze-out is projected on the assumed 
dilambda wave
function (harmonic oscillator) via the Wigner-function method as described
in \cite{rafi}.
According to the weak coupling between $\Lambda$'s in
mean-field calculations \cite{Sch92} we assume the same coupling for
$\Lambda\Lambda$-cluster as for deuterons\cite{rafi}.

In Fig.~\ref{yclust} the calculated rapidity distributions of hyperons and
$\Lambda\Lambda$-clusters are shown for central reactions of heavy systems 
at AGS and SPS energies. The multiplicities of $\Lambda$'s plus $\Sigma^0$'s
in inelastic p+p reactions are $0.088\pm 0.003$ at 14.6~GeV/c and $0.234\pm
0.005$ at 200~GeV/c with the present version of the model. These numbers are
given to assess the absolute yields in A+A collisions. The hyperon 
rapidity density stays almost constant when going from AGS to SPS energies,
the ${\rm d}N/{\rm d}y$ of the hyper-clusters even drops slightly at
midrapidity. This is due to the higher temperature which gives rise to
higher relative momenta and therefore a reduced cluster probability.
The $\Lambda/\Lambda\Lambda$ ratio is approximately 100, which can be
compared to the $\Lambda/H^0$ ratios which result from the expanding
quark gluon plasma (see last section).

%

\end{document}